\documentclass[conference]{IEEEtran}
\IEEEoverridecommandlockouts

\usepackage{cite}
\usepackage[cmex10]{amsmath}
\usepackage{algorithmic}
\usepackage{array}
\usepackage{url}
\usepackage{footnote}
\usepackage{float}
\usepackage[utf8]{inputenc}

\usepackage[T1]{fontenc}
\usepackage[ruled,vlined]{algorithm2e}
\usepackage{ifthen}
\usepackage{graphicx}
\usepackage{comment}
\usepackage{amsthm}
\usepackage{amsmath}
\usepackage{amsmath,amssymb,amsfonts}
\usepackage{nicefrac}
\usepackage{algorithmic}
\usepackage{bm}
\usepackage{multirow}
\usepackage{mathtools}
\usepackage{tikz}
\usetikzlibrary{calc,math}
\usepackage[framemethod=TikZ]{mdframed}
\usepackage{pgfplots}
\pgfplotsset{compat=1.16}
\usepackage{hyperref}
\hypersetup{colorlinks=true, unicode=true, linkcolor=[rgb]{0.10,0.05,0.67}, citecolor=[rgb]{0.10,0.05,0.67}, filecolor=[rgb]{0.10,0.05,0.67}, urlcolor=[rgb]{0.10,0.05,0.67}}

\newtheorem{theorem}{Theorem}

\newtheorem{corollary}{Corollary}
\newtheorem{example}{Example}

\definecolor{DarkGreen}{rgb}{0.1,0.5,0.1}
\definecolor{DarkRed}{rgb}{0.5,0.1,0.1}
\definecolor{DarkBlue}{rgb}{0.1,0.1,0.5}
\definecolor{DarkPurple}{rgb}{0.5,0.2,0.5}
\definecolor{DarkTurquoise}{rgb}{0.1,0.5,0.5}

\makeatletter
\renewcommand*{\eqref}[1]{%
  \hyperref[{#1}]{\textup{\tagform@{\ref*{#1}}}}%
}

\makeatother

\begin{document}

\title{On the Benefits of Coding for Network Slicing\\
\thanks{This is a slightly modified version of a paper that will be presented at the IEEE ICC 2024. This work is supported by the SNOB-5G project under the MIT Portugal Program and by JMA Wireless.}}

\author{
\IEEEauthorblockN{Homa Esfahanizadeh$^{\dagger}$, Vipindev Adat Vasudevan$^{\ddagger}$, Benjamin D. Kim$^{+}$,  Shruti Siva$^{\ddagger}$, Jennifer Kim$^{\ddagger}$,\\Alejandro Cohen$^{*}$, and Muriel M\'edard$^\ddagger$}
\IEEEauthorblockA{
$^\dagger$Nokia Bell Labs, Murray Hill, NJ, USA, Email: homa.esfahanizadeh@nokia-bell-labs.com
}
\IEEEauthorblockA{
$^\ddagger$Massachusetts Institute of Technology (MIT), Cambridge, USA, Emails: \{vipindev, shrutsiv, jennkim, medard\}@mit.edu
}
\IEEEauthorblockA{
$^+$University of Illinois Urbana-Champaign (UIUC), Champaign, Illinois, USA, Email: bdkim4@illinois.edu
}
\IEEEauthorblockA{
$^*$Technion—Israel Institute of Technology, Haifa, Israel, Email: alecohen@technion.ac.il
}
}

\maketitle

\begin{abstract}
Network slicing has emerged as an integral concept in 5G, aiming to partition the physical network infrastructure into isolated slices, customized for specific applications. We theoretically formulate the key performance metrics of an application, in terms of goodput and delivery delay, at a cost of network resources in terms of bandwidth. We explore an un-coded communication protocol that uses feedback-based repetitions, and a coded protocol, implementing random linear network coding and using coding-aware acknowledgments. We find that coding reduces the resource demands of a slice to meet the requirements for an application, thereby serving more applications efficiently. Coded slices thus free up resources for other slices, be they coded or not. Based on these results, we propose a hybrid approach, wherein coding is introduced selectively in certain network slices. This approach not only facilitates a smoother transition from un-coded systems to coded systems but also reduces costs across all slices. Theoretical findings in this paper are validated and expanded upon through real-time simulations of the network.
\end{abstract}

\section{Introduction}


With wireless technologies evolving to the 5-th generation (5G) and beyond, the standardization bodies like the International Telecommunication Union (ITU) and 3GPP have identified different use cases that drive the innovations \cite{3gpp.22.861,3gpp.22.862,3gpp.22.863}. The three broad families of use cases in 5G are enhanced mobile broadband (eMBB), ultra-reliable low latency communications (URLLC), and massive machine-type communications (mMTC). While the eMBB use cases prioritize achieving high data rates and traffic density, the URLLC focuses on serving critical applications with low latency and high reliability. The large number of connected devices and the heterogeneous nature of devices are characterized by the mMTC use case.

\begin{figure}
    \centering
    \includegraphics[trim=0cm 0.0cm 0cm -0.2cm,width = 1\columnwidth]{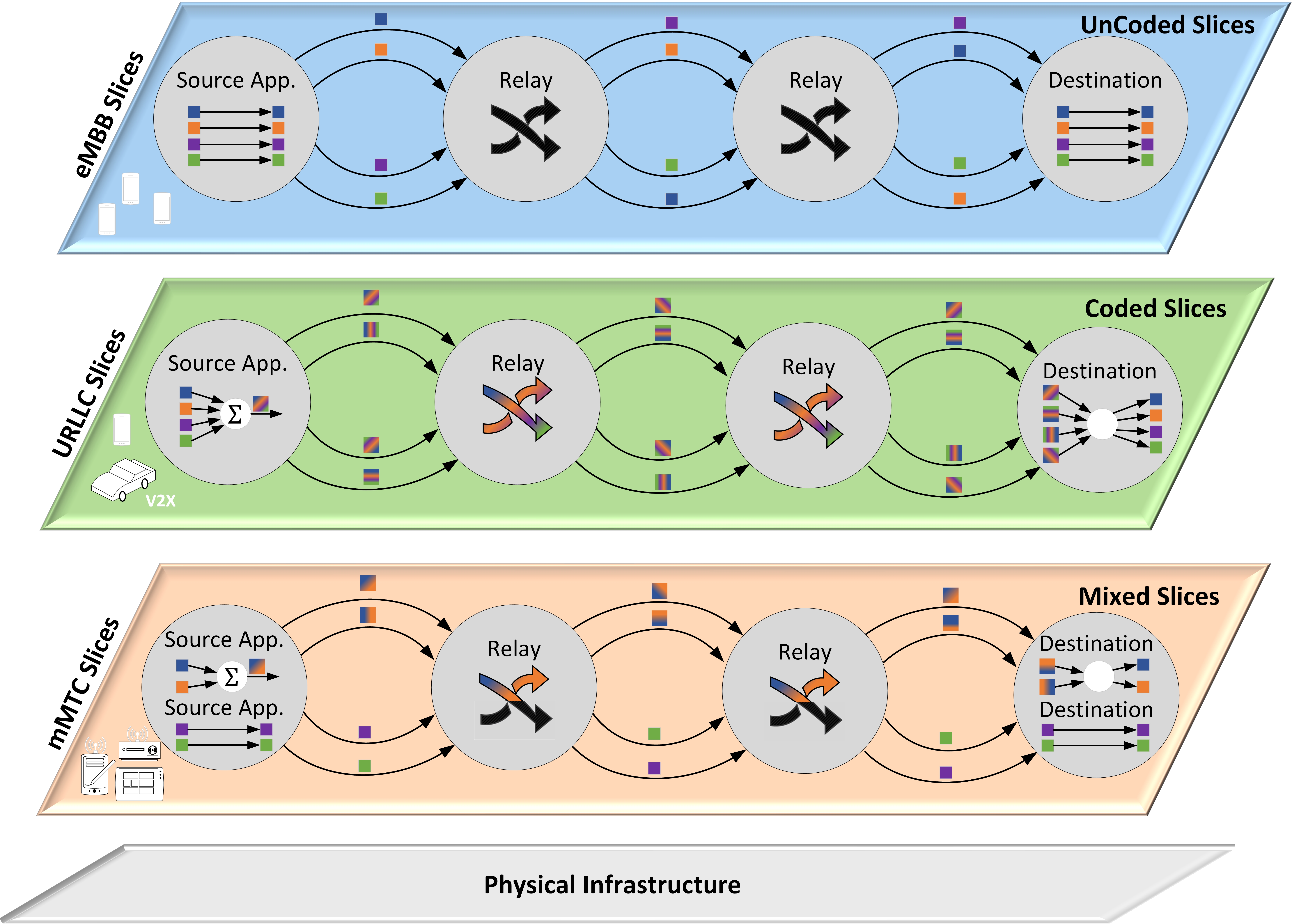}
    \caption{Heterogeneity in advanced meshed communications with hybrid technologies. The slices, which share the same network infrastructure, can use different communication protocols, and they serve different 5G use cases.\vspace{-0.2cm}}
    \label{fig:mixed_slices}
    \vspace{-0.2cm}
\end{figure}

Evident from the broad range of applications with different requirements in the upcoming communication networks, there is no single model that can optimally serve all of them. Thus, several novel concepts have emerged to define more advanced and efficient networks that can satisfy a wide range of user requirements. Among them is \textit{network slicing}, which is a mechanism for optimally allocating network resources to several services and applications that have different requirements \cite{alliance2016description,ordonez2017network}. Slicing has become an important concern in the planning and deployment of the network architectures and is expected to continue and grow as an integral part of 6G networks \cite{zhang2019overview,wu2022ai}. Network slicing in the future generations of communication networks will enable various applications and services to jointly and efficiently benefit from the shared network resources, via defining several virtual data pipelines (e.g., \textit{virtual networks} \cite{cohen2021bringing}). Emerging applications demand strong guarantees on their performance metrics, such as latency and bandwidth, only getting stronger and more challenging to be satisfied over time. Meeting these stringent requirements in a dense and heterogeneous network is often not trivial \cite{joshi2012playback,dias2023sliding}, demanding efficient and real-time network slicing strategies for defining virtual networks, which is the scope of this paper.

Different slices may employ different communication protocols to comply with the latest technology and demands of their associated applications. For instance, in an eMBB use case with a focus on the bandwidth, the protocols that achieve the maximum transmission rate (such as Automatic Repeat Request (ARQ) \cite{anagnostou1986performance}) are preferred, while in the URLLC use case, advanced coding schemes can be deployed to achieve higher reliability and low latency communications. When it comes to the massive IoT use case, the requirements of different applications in the IoT environment can be very different from each other, and the participating devices may have different computational capabilities. Thus, in an mMTC use case, multiple transmission techniques (coded or un-coded) can be deployed. Fig.~\ref{fig:mixed_slices} shows how slicing can be used to serve the diverse 5G use cases.

Network coding (NC) is a communication solution, which can significantly improve the communication quality over lossy channels on several fronts, e.g., reliability and latency. NC was first proposed as a methodology to achieve the capacity of the network \cite{ahlswede2000network}, and since then, it has been extended and equipped with several techniques \cite{ho2006random,cohen2020adaptive,Vasudevan2023Practical} to offer reliable high-bandwidth communication with low latency and cost, making it a great candidate for future wireless scenarios \cite{cohen2021adaptive,michel2022flec,cohen2022broadcast,enenche2023network}. Coding over network slices can significantly enhance service quality and ensure that the requirements are satisfied with the available resources \cite{cohen2021bringing,fitzek2020computing,10187820}. Network coding implementations have been proven to perform better and coexist in a network competing for resources without degrading the performance of non-coding data flows \cite{kim2014congestion}. 

In this paper, we build upon existing network coding innovations and explore the multitude of benefits they provide in network slices. We show the benefits of network coding for virtual network slicing, where coding may be applied to a fraction of the network slices. We focus primarily on the eMBB and URLLC use cases, where high throughput and low latency are the priorities, respectively. Further, we discuss the scenarios of dense heterogeneous networks, which can be associated with the mMTC use case and may have parts with and without network coding capabilities. We also propose dynamic slicing techniques where allocated resources can be released and reallocated to other applications/slices, once a particular slice finishes serving its application.


\section{System Model and Problem Setting}\label{sec:system_model}

Consider a multi-path network that consists of $n$ parallel binary erasure channels (BECs) with erasure probabilities in $\mathcal{P}=\{p_1,\dots,p_n\}$. Such a model complies with the concept of virtual networks, where global paths are identified and their end-to-end erasure rates are estimated \cite{cohen2021bringing}. At each time slot, one packet is transmitted through each link, which is lost with a probability advised by the link parameter. \textit{Network slicing} is a partitioning of the network links into $J$ slices, where the links in $\mathcal{P}_j\subseteq\mathcal{P}$ are assigned to the $j$-th slice and $\cup_{j=1}^J \mathcal{P}_j=\mathcal{P}$. Applications that share the network infrastructure have different requirements, and the questions we discuss in this paper are how to characterize the performance of an application given its slice bandwidth and its communication protocol, and how to split the resources into several slices considering diverse requirements of each application.

We first define the key performance metrics of an application that acquires the $j$-th slice, $j\in\{1,\dots,J\}$. The performance metrics depend on resources that are allocated to the application, i.e., $\mathcal{P}_j$, as well as the deployed communication solution by the application, e.g., random linear network coding (RLNC) or selective repeat (SR)- ARQ.
\begin{itemize}
    \item \textbf{Delivery Delay (resp., In-Order Delivery Delay):} Number of time slots it takes for an information packet to be delivered (resp., delivered in-order) at the destination, denoted with {$D(\mathcal{P}_j)$} (resp., {$I(\mathcal{P}_j)$}).
    \item \textbf{Goodput:} Number of information packets that are delivered per time slot, denoted with {$G(\mathcal{P}_j)$}.
    \item \textbf{Completion Time:} Number of time slots it takes for an application to successfully deliver $\nu$ information packets, denoted with {$T_\nu(\mathcal{P}_j)$}.
\end{itemize}

Ideally, we need to slice the network such that different requirements of each application are satisfied, e.g., $E[G(\mathcal{P}_j)]\geq \bm{G}_j$ and $E[D(\mathcal{P}_j)]\leq \bm{D}_j$, where $\bm{G}_j$ is the minimum demanded \textit{goodput} and $\bm{D}_j$ is the maximum tolerable \textit{delivery delay} for the $j$-th application, so its quality of service will not be degraded. 

\section{Theoretical Characterization}\label{sec:theory}

In this section, we characterize the performance of an application given its slice bandwidth and its communication protocol. As our benchmarks, we explore two communication protocols, i.e., SR-ARQ  as a representative of an un-coded communication solution and RLNC  as a representative of a coded solution. The considered communication system is time-slotted, and an acknowledgment (ACK/NACK) is received per transmitted packet after one round trip time ($RTT$), corresponding to the communication in the transport layer of a network. For simplicity of the theoretical analysis, we assume $RTT$ is the same for all links, and we leave the multi-path scenario with heterogeneous $RTT$ values for future work.

\subsection{Communication Protocol 1 (Un-coded):  SR-ARQ}\label{sec:uncoded}

In this setting, the transmitted packets are un-coded. When a packet is sent, its acknowledgment is received (ACK or NACK), after $RTT$ time slots. Per receipt of a NACK for a packet, the sender re-transmits the same packet in the next time slot through a randomly-chosen link of the slice.\footnote{We assume the sender's buffer does not have size limitations.}

According to the definition, \textit{delivery delay} of a packet for the $j$-th application $D(\mathcal{P}_j)$ is the difference between the time the packet is transmitted for the first time through its allocated slice, until the time it is successfully received. We assume the transmission time of a packet (including the first transmission and the possible re-transmissions) is negligible compared to $RTT$. Thus, 
\begin{equation*}
    D(\mathcal{P}_j)\in\left\{\frac{RTT}{2},\frac{RTT}{2}+RTT,\frac{ RTT}{2}+2RTT\dots\right\}.\vspace{-0.1cm}
\end{equation*}
If there was no error, the goodput was $G(\mathcal{P}_j)=|\mathcal{P}_j|$ (i.e., same as throughput). However, because of packet loss and re-transmissions, the goodput is typically less than this bound.

\begin{theorem}
\label{theorem:avg_var_DandG_MP_SR_ARQ}
Average delivery delay and average goodput for the multi-path SR-ARQ communication solution are
\begin{equation} \label{eq:AVG_D_G_ARQ}  
\begin{split}
E[D(\mathcal{P}_j)]&=0.5\frac{1+\overline{\mathcal{P}_j}}{1-\overline{\mathcal{P}_j}}RTT, \\
E[G(\mathcal{P}_j)]&=\sum_{p_i\in\mathcal{P}_j} (1-p_i)=|\mathcal{P}_j|(1-\overline{\mathcal{P}_j}).
\end{split}
\end{equation}
Here, $\overline{\mathcal{P}_j}=\left(\sum_{p_i\in\mathcal{P}_j} p_i\right)/{|\mathcal{P}_j|}$ is the average erasure probability of the slice.
\end{theorem}

\begin{proof}
    We first show that the distribution of delivery delay is,
    \begin{equation}
        \label{eq:PMF_D_ARQ}P\left[D(\mathcal{P}_j)=\frac{RTT}{2}+kRTT\right]=
            \overline{\mathcal{P}_j}^k\left(1-\overline{\mathcal{P}_j}\right),
    \end{equation}
    when $k\in\{0,1,\dots\}$, and zero otherwise.  We prove this by induction. We first show the result holds for $k=0$:
\begin{equation*}
    \begin{split}
        P\Bigg[D&(\mathcal{P}_j)=\frac{RTT}{2}\Bigg]\\
        &=\sum_{p_i\in\mathcal{P}_j}\frac{1}{|\mathcal{P}_j|}P\Bigg[D(\mathcal{P}_j)=\frac{RTT}{2}|\text{$i$-th link is selected}\Bigg]\\
        &=\frac{1}{|\mathcal{P}_j|}\sum_{p_i\in\mathcal{P}_j} (1-p_i)=(1-\overline{\mathcal{P}_j}).
    \end{split}
    \end{equation*}
    Assume $P[D(\mathcal{P}_j)=\frac{RTT}{2}+kRTT]=\overline{\mathcal{P}_j}^k(1-\overline{\mathcal{P}_j})$. Then,
    \begin{equation*}
        \small
        \begin{split}
            &P\left[D(\mathcal{P}_j)=\frac{RTT}{2}+(k+1)RTT\right]=\\
            &=\sum_{p_i\in\mathcal{P}_j}\hspace{-0.2cm}\frac{1}{|\mathcal{P}_j|}P\left[D(\mathcal{P}_j){=}\frac{RTT}{2}+(k+1)RTT|\text{$i$-th link is selected}\right]\\
            &=\frac{1}{|\mathcal{P}_j|}\sum_{p_i\in\mathcal{P}_j} p_i P\left[D(\mathcal{P}_j)=\frac{RTT}{2}+kRTT\right]\\
            &=\overline{\mathcal{P}_j}^k(1-\overline{\mathcal{P}_j})\frac{1}{|\mathcal{P}_j|}\sum_{p_i\in\mathcal{P}_j} p_i=\overline{\mathcal{P}_j}^{k+1}(1-\overline{\mathcal{P}_j}),
    \end{split}
\end{equation*}
and this completes the proof, by induction, for the PMF expression in (\ref{eq:PMF_D_ARQ}). Next, we derive the expectation,
\begin{equation*}
    \begin{split}
        E[D(\mathcal{P}_j)]&=\sum_{k=0}^{\infty}\left(\frac{RTT}{2}+kRTT\right)\overline{\mathcal{P}_j}^{k}(1-\overline{\mathcal{P}_j})\\
        &=\frac{RTT}{2}+RTT(1-\overline{\mathcal{P}_j})\sum_{k=0}^{\infty}k\overline{\mathcal{P}_j}^{k}\\
        &=\frac{RTT}{2}+RTT(1-\overline{\mathcal{P}_j})\frac{\overline{\mathcal{P}_j}}{(1-\overline{\mathcal{P}_j})^2}\\
        &=0.5 \frac{1+\overline{\mathcal{P}_j}}{1-\overline{\mathcal{P}_j}}RTT.
    \end{split}
\end{equation*}

We now prove the expression for the average goodput. Let $X_i$ be a binary random variable that takes value one if the $i$-th link has a successful transmission, and zero otherwise. Thus, $E[X_i]=1-p_i$. By definition, $G(\mathcal{P}_j)=\sum_{p_i\in\mathcal{P}_j}X_i$. Because of linearity of the expectation, we have
\begin{equation*}
    E[G(\mathcal{P}_j)]=\sum_{p_i\in\mathcal{P}_j}E[X_i]=\sum_{p_i\in\mathcal{P}_j}(1-p_i)=|\mathcal{P}_j|(1-\overline{\mathcal{P}_j}).
\end{equation*}
\end{proof}

As we see, the average delivery delay for SR-ARQ only depends on the average erasure probability of the links in the slice. Therefore, a slice with only one reliable link is preferred to another slice with several mediocre links for this communication protocol, in terms of the delivery delay. However, the average goodput cannot decrease when an arbitrary link is added to the slice. 

\subsection{Communication Protocol 2 (Coded): RLNC}

This communication solution is an adaptive RLNC scheme. Consider the $j$-th application with the allocated slice $\mathcal{P}_j$, $j\in\{1,\dots,J\}$. A \textit{generation} is defined as a sequence of original packets that are coded together; the generation size is denoted with $k_j\gg |\mathcal{P}_j|$. The sender sequentially transmits the coded generations, starting with the first generation. For each generation, it transmits $k_j\gamma_j^1$ coded packets through the links in $\mathcal{P}_j$. The parameter $\gamma_j^1$ is called the forward error correction (FEC) rate, and a reasonable choice is proportional to $1/(1-\overline{\mathcal{P}_j})$. Upon receiving feedback of the last-sent encoded packet of a generation, the sender knows that the receiver requires ${m}_j$ more coded packets (i.e., missing degrees of freedom) to decode the generation. It then transmits $\gamma_j^2 {m}_j$ coded packets in the next time slot. The parameter $\gamma_j^2$ is called the feedback-based (FB) rate, and is set such that the probability that the receiver fails to decode the generation after the second round is almost zero. If the sender has the option to select among the paths, its strategy is a random selection. With this setting, we ensure that all packets are transmitted successfully, meeting the ultra-reliable communication requirements.

We first characterize the random variable ${m}_j$. The probability of ${m}_j = 0$ is equal to having equal or less than $\lceil(\gamma_j^1-1)k_j\rceil$ failed transmissions in the first trial of a generation. Similarly, the probability of ${m}_j = m$, $0<m\leq k_j$, is equal to having  $\lceil(\gamma_j^1-1)k_j\rceil + m$ failed transmissions in the first trial. 

\begin{theorem}\label{theorem:prob_missing_NC}
Distribution of the random variable $m_j$ is approximated as follows,
    \begin{equation}
    \label{dist_m}
    P[{m_j}=m]\approx\begin{cases}
        \sum_{f=0}^{\left\lceil\left(\gamma_j^1-1\right)k_j\right\rceil}\frac{\lambda^{f}e^{-\lambda}}{f!}&m=0,\\\vspace{-0.2cm}\\
        \frac{\lambda^{\left\lceil\left(\gamma_j^1-1\right)k_j\right\rceil + m}e^{-\lambda}}{\left(\left\lceil\left(\gamma_j^1-1\right)k_j\right\rceil+m\right)!}&m\in\{1,\dots,k_j\},\\\vspace{-0.2cm}\\
        0&\text{otherwise},
    \end{cases}
\end{equation}
where $\lambda=k_j\gamma_j^1\overline{\mathcal{P}_j}$.
\end{theorem}
\begin{proof}
The number of failures in a trial can be modeled as summation of independent non-identical Bernoulli random variables, assuming the links are memoryless. If the erasure probabilities (Bernoulli parameters) are close to zero, this distribution can be approximated by a Poisson distribution with its parameter being the summation of the parameters of the Bernoulli random variables \cite{10.1214/aoms/1177705799}, i.e., 
$$\lambda=\frac{k_j\gamma_j^1}{|\mathcal{P}_j|}\sum_{p_i\in\mathcal{P}_j} p_i=k_j\gamma_j^1\overline{\mathcal{P}_j}.\vspace{-0.5cm}$$
\end{proof}
\begin{corollary}
    The average delivery delay for the presented multi-path RLNC communication solution is,
    \begin{equation}
        \label{eq:AVG_D_RLNC}
        \begin{split}
            E[D(\mathcal{P}_j)]&\approx\left(\frac{RTT}{2}+\left\lceil{\frac{k_j\gamma_j^1}{|\mathcal{P}_j|}}\right\rceil\right)P[{m_j}=0]\\
            &+\left(\frac{3RTT}{2}+\left\lceil{\frac{k_j\gamma_j^1}{|\mathcal{P}_j|}}\right\rceil+1\right)(1-P[{m_j}=0]).
    \end{split}
\end{equation}
\end{corollary}
\begin{proof}
    The distribution of delivery delay for the presented RLNC solution is,\footnote{In this analysis, we assume the decoding delay is negligible.}
\begin{equation*}
    D(\mathcal{P}_j)\approx
    \begin{cases}
        \frac{RTT}{2}+\left\lceil{\frac{k_j\gamma_j^1}{|\mathcal{P}_j|}}\right\rceil,& {m_j}=0,\\\vspace{-0.2cm} \\
        \frac{3 RTT}{2}+\left\lceil{\frac{k_j\gamma_j^1}{|\mathcal{P}_j|}}\right\rceil+1,& {m_j}>0.
    \end{cases}
\end{equation*}
Therefore, the average delivery delay for RLNC can be obtained as in (\ref{eq:AVG_D_RLNC}).
\end{proof}

\begin{corollary}
The average goodput for the presented multi-path RLNC communication solution is,
\begin{equation}
\label{eq:AVG_G_RLNC}
    \begin{split}
        E[G(\mathcal{P}_j)]&=E\left[ \frac{k_j}{k_j\gamma_j^1+{m_j}\gamma_j^2}|\mathcal{P}_j|\right]=\\&\sum_{m=0}^{k_j} \frac{k_j}{k_j\gamma_j^1+m\gamma_j^2}P[{m_j}=m]|\mathcal{P}_j|.
    \end{split}
\end{equation}
\end{corollary}

We finish this subsection by presenting examples that demonstrate the performance of network slices that use the coded solution compared to those that use the un-coded one, for utilizing the shared network infrastructure.

\begin{example} \label{example:Hetro}
    Consider a network with $10$ links and $RTT=1000$, modeling a high-rate transmission scenario. The erasure probabilities are specified by the vector $\mathcal{P}$ given below, where the $i$-th element is the erasure probability of the $i$-th link:
    \begin{equation*}
        \mathcal{P}=[0.05,0.01,0.08,0.02,0.06,0.01,0.07,0.09,0.09,0.01].
    \end{equation*}
    We evaluate the delivery delay of two slices that share this network. We consider the $i$-th slicing choice as $\mathcal{P}_1=\{p_1,\dots,p_i\}$ and $\mathcal{P}_2=\{p_{i+1},\dots,p_{10}\}$. The un-coded protocol is SR-ARQ, and the coded protocol is RLNC with parameters $k_j=50$, $\gamma_j^1=1.1/(1-\overline{\mathcal{P}_j})$, and $\gamma_j^1=2/(1-\overline{\mathcal{P}_j})$, for $j\in\{1,2\}$.\footnote{All theoretical results are validated using Monte Carlo simulations.} 

    Fig.~\ref{fig:analysis_Hetro} shows the average delivery delay (left) and average goodput (right) for various slicing choices and communication protocols. We observe that the delay of an application that uses coded communication protocol is consistently lower than the one that uses un-coded solution, across all slicing choices. The difference is more significant when the size of the slice grows, and such a benefit comes at slight cost of a small degradation in the average goodput.

    \begin{figure}
    \centering
    \hspace{-0.3cm}\includegraphics[width=0.48\textwidth]{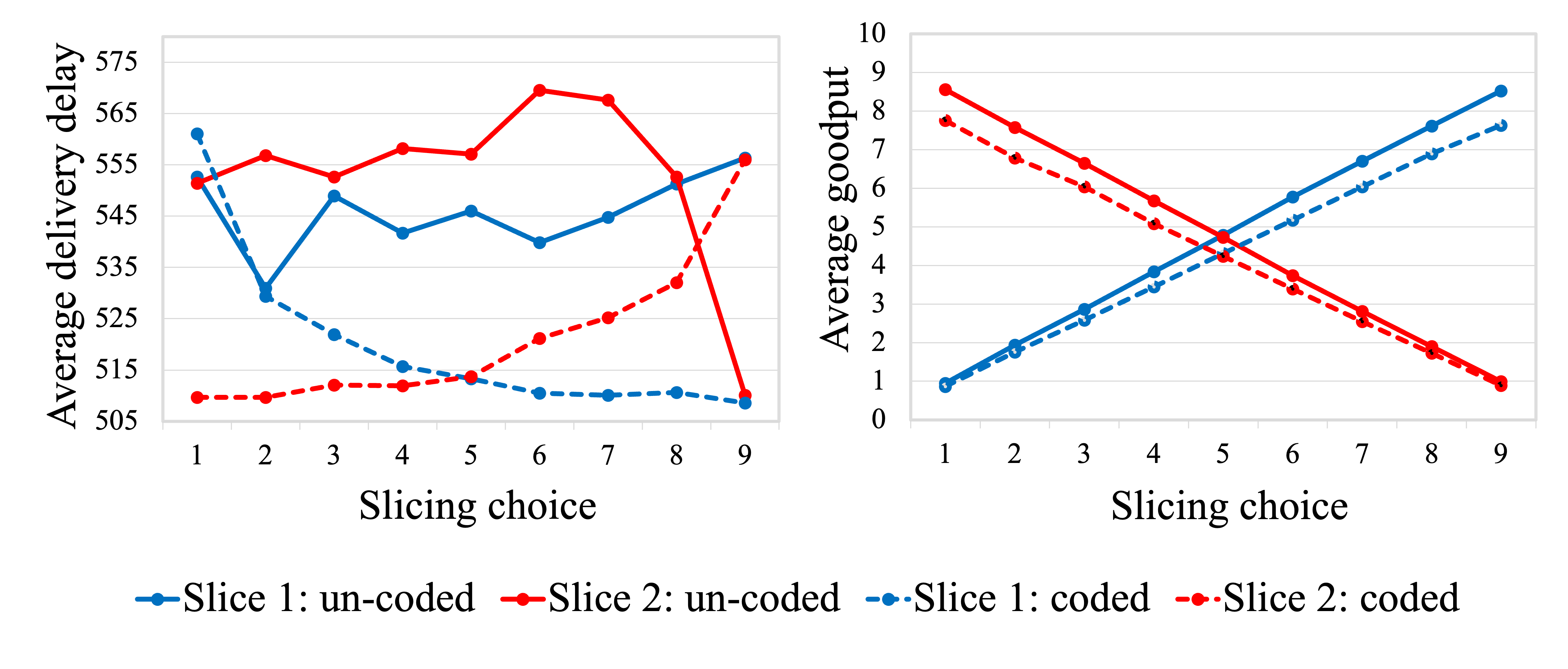}\vspace{-0.4cm}
    \caption{Average delivery delay and goodput for different slicing choices and communication protocols, over a network with $10$ heterogeneous links.\vspace{-0.5cm}}
    \label{fig:analysis_Hetro}
    \end{figure}
\end{example}

\subsection{Communication-Aware Network Slicing}

Here, we study networks with a mixed collection of slices, where some of them are coded and some are un-coded. We manifest the benefits of using coded slices for the whole network. By using coding, applications need smaller slices, which not only reduces the cost for that application but also frees up more resources for the other applications that may need more resources. Thus, the same physical network can be shared by more delay-sensitive applications, when the coded communication protocol is used by the applications. This mutual benefit is particularly important for 5G uses cases that have various requirements. For instance, in the mMTC, the network needs to be shared with many applications with potentially different requirements, and it is critical to consider their differences to optimize the network to serve more and more applications. Our result motivates incorporating the statistical measures of communication protocols given their assigned bandwidth into the network slicing strategy. 

The next example highlights the importance of using coding in realizing more economical networks, as it shows the difference between the number of delay-sensitive applications that can be served in a network.

\begin{example}
    Consider a network with $10^4$ BEC links, each with erasure probability $0.1$, and $RTT=1000$. The goal is to serve as many delay-sensitive applications (e.g., URLLC) as possible over this network, with and without coding. The requirement of each application is to have a delivery delay of less than $530$ time slots, and to have goodput of at least $5$ packets per time slot.

    We first consider the un-coded communication protocol. In order to satisfy the goodput requirement, each application needs to have at least $6$ links, as derived below, see Theorem~\ref{theorem:avg_var_DandG_MP_SR_ARQ},
    \vspace{-0.2cm}
    \begin{equation*}
        E[G(\mathcal{P}_j)]\geq 5 \rightarrow
            |\mathcal{P}_j|(1-\overline{\mathcal{P}_j})\geq 5,\rightarrow
            |\mathcal{P}_j|\geq 5.56.
    \end{equation*}
    Based on the theoretical derivations in Theorem~\ref{theorem:avg_var_DandG_MP_SR_ARQ}, the average delivery delay of an un-coded slice is only dependent on the average erasure probability, and it is equal to $611.11$ time slots for the current network configuration, regardless of the size of the slice. As seen, the delay requirement cannot be satisfied, regardless of the slicing choice, if the un-coded solution is deployed by the applications.

    Next, we consider the coded communication protocol with $k_j=50$, $\gamma_j^1=1.1/(1-\overline{\mathcal{P}_j})=1.22$, and $\gamma_j^1=2/(1-\overline{\mathcal{P}_j})=2.22$, for $j\in\{1,\dots,J\}$. Based on Theorem~\ref{theorem:prob_missing_NC}, we have $\lambda=k_j\gamma_j^1\overline{\mathcal{P}_j}=6.1$, and $P[{m_j}=m] = [0.9776, 0.0124, 0.0058, 0.0025, 0.0010, 0.0004]$, for $m = 0,1,2,3,4,5$, respectively, and negligible for $m > 5$.
    
    To satisfy the delay requirement, each application needs to have at least $9$ links, as derived below, see (\ref{eq:AVG_D_RLNC}),
        \begin{equation*}
        \begin{split}
            &E[D(\mathcal{P}_j)]\leq 530 \\
            &\left(500+\left\lceil{\frac{61}{|\mathcal{P}_j|}}\right\rceil\right)0.9776+\left(1501+\left\lceil{\frac{61}{|\mathcal{P}_j|}}\right\rceil\right)0.0224\leq 530
        \end{split}
        \end{equation*}     
        \begin{equation*}
        \begin{split}
            &\left\lceil{\frac{61}{|\mathcal{P}_j|}}\right\rceil\leq7.5776\rightarrow |\mathcal{P}_j|\geq 8.05.
        \end{split}
        \end{equation*}
    
    To satisfy the goodput requirement, each application needs to have at least $7$ links, derived as follows, see (\ref{eq:AVG_G_RLNC}):
    \begin{equation*}
    \begin{split}
        &E[G(\mathcal{P}_j)]\geq 5 \rightarrow
        \sum_{m=0}^{5} \frac{50}{61+2.22 m}P[{m_j}=m]|\mathcal{P}_j|\geq 5\\
        &\rightarrow
        0.8184|\mathcal{P}_j|\geq 5\rightarrow |\mathcal{P}_j|\geq 6.109.
        \end{split}
    \end{equation*}

Therefore, each application needs to acquire at least $9$ links to meet its requirements, and the network can potentially serve 111 such applications. 
    
\end{example}

\section{Real-Time Simulation of a Network with Mixed Slices}\label{sec:simulations}

In this section, we expand upon our theoretical results on the network with mixed slices. In particular, we explore our implementations of both coded slices (using RLNC) and un-coded slices (using SR-ARQ) through SimPy, a discrete-event simulator \cite{simpy}. SimPy excels in simulating real-world networking scenarios thanks to its adeptness in handling asynchronous events, time-dependent behaviors, and custom event scheduling. By adopting a process-based paradigm, SimPy enabled us to effectively simulate data flow within the network and replicate desired client-server transactions. 

For our experiments, we consider a homogeneous network with the total number of $n=20$ links and $RTT=150$ slots, where each link has an erasure probability of $0.2$. The network consists of two slices serving two different applications (each may have varying requirements). The first slice is dedicated to the first application, and is allocated $i$ out of the total available links, $i\in\{1,\dots,20\}$. The parameter $i$ indicates the slicing choice, and the second slice serves the second application with the remaining resources.

\begin{figure}
    \centering
    \begin{tabular}{cc}
    \includegraphics[trim=1.0cm 0cm -0.6cm 0cm,clip,width=0.50\textwidth]{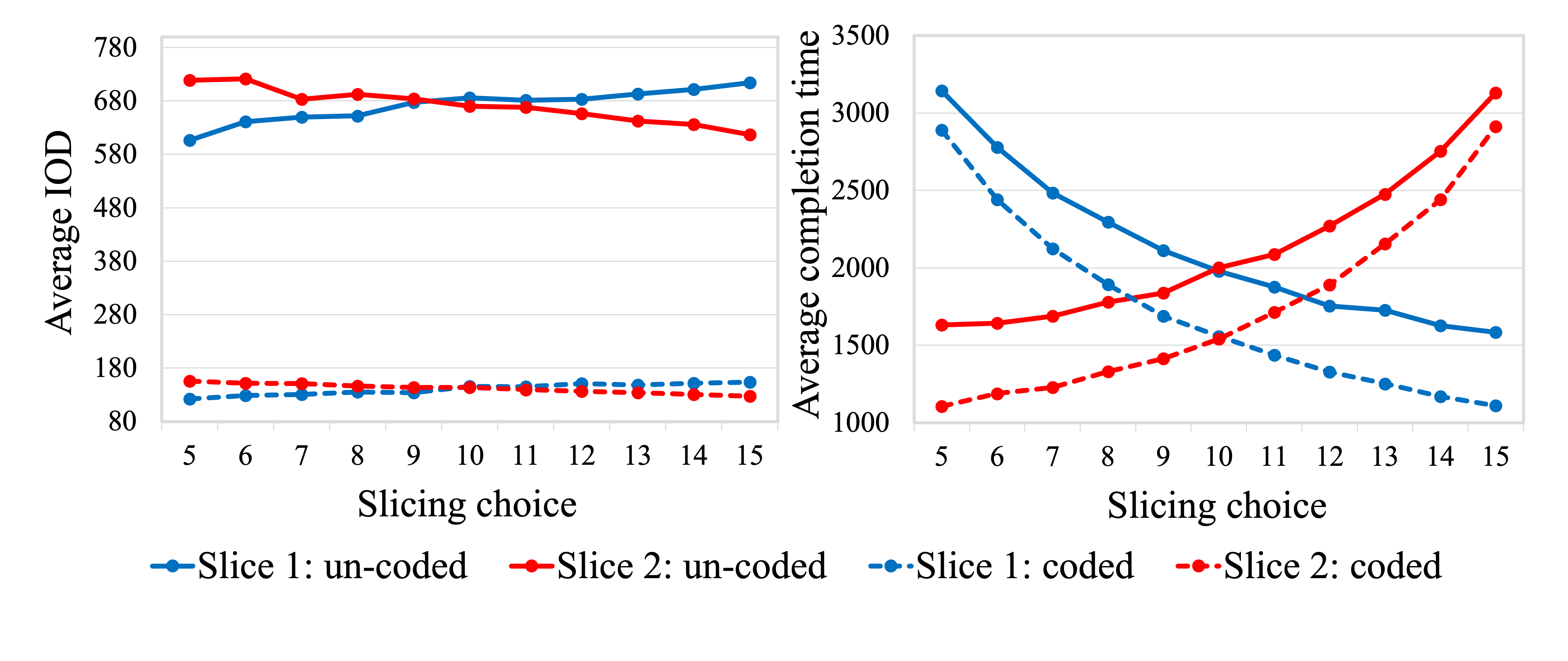}& 
    \end{tabular}
    \vspace{-0.6cm}
    \caption{(left): Average in-order delivery delay (IOD); (right): average completion time, for different slicing choices and communication solutions.\vspace{-0.5cm}}
    \label{fig:exp_1_2}
\end{figure}

Fig.~\ref{fig:exp_1_2}~(left) shows that the average in-order delivery delay slightly increases with the number of links allocated to a slice. When a larger slice is available, more packets will be sent at the same time slot. Then the loss of a single packet impacts more packets in transit, increasing the in-order delivery delay. However, the RLNC scheme results in significantly lower in-order delivery delay since it sends additional repair packets apriori to compensate for any losses in advance. Even though the average in-order delivery delay increases with number of available channels, the completion of both coded or un-coded slices decrease with it, since more transmitted packets at each time slot results in a faster completion for a specific application, as shown in Fig.~\ref{fig:exp_1_2}~(right).


RLNC helps the coded slice to achieve a lower average delay and a significantly lower in-order delay. In fact, the coded slice can meet the low-delay requirements with a smaller slice as well, leaving more resources for other applications that require more bandwidth (e.g., eMBB slice). The following experiments demonstrate this scenario considering two applications with different requirements. We assume each time slot in our network model is  $50~\mu sec$, and each link has a bandwidth of $28$~Mbps (i.e., each link can send a packet of $1400$ bits per time slot). 

The first application is a delay-sensitive application, such as tactile internet, that requires an average delay of less than $5ms$ and throughput of at least $10$~Mbps (as per \cite{3gpp.22.862,alliance20155g}). The second application is a more throughput-demanding application, such as an 8K 3D video streaming, that requires at least $250$~Mbps with a delay tolerance of around $10${-}$12$ ms\cite{3gpp.22.863,alliance20155g}\footnote{The throughput requirements in the standards can directly be considered as goodput requirements in our settings.}. In our experimental setting, 5 ms corresponds to $100$ time slots, and a goodput of $250$~Mbps can be obtained if at least $8.92$ transmissions per slot occur. The first application requires $E[I(\mathcal{P}_1)]\leq 100$ time slots (modeling URLLC application), while the second application requires $E[G(\mathcal{P}_2)]\geq 9$ packets per time slot (modeling an eMBB slice). 


The requirements of the first application can never be met with an un-coded communication protocol. Introducing coding in the first slice will make it possible to realize the delay requirement of application 1, even with a very small number of links. This provides the second slice with enough resources to support the high goodput requirements of application 2, enabling more feasible slicing choices, as shown in table \ref{tab_exp_2}. Furthermore, the in-order delay is significantly lower within coded slices assuring seamless communication. Thus, the mixed slices satisfy both application requirements with the same amount of resources, while the un-coded slices fail to achieve at least one set of requirements. This shows that the coded slice is essential to achieve ultra-reliable low-latency requirements in a practical setting. Further, the coded slice can achieve this performance with a smaller slice of resources as well, making room for more applications to be served with the same physical infrastructure. 

\begin{table*}
    \centering
    \caption{This table shows the performance of un-coded slices and mixed slices settings. The \textcolor{purple}{\textbf{purple}} colored entries show the slices that match the eMBB requirements, while those with \textcolor{blue}{\textbf{blue}} color show the URLLC slices}
    \vspace{-0.2cm}
    \label{tab_exp_2}
    \begin{tabular}{|c||c|c|c|c|c|c||c|c|c|c|c|c|}
    \hline
    \multirow{2}{*}{Slicing Choice}&\multicolumn{6}{c||}{Un-coded slices}&\multicolumn{6}{c|}{Mixed Slicing} \\ \cline{2-13}
    ~ & \multicolumn{3}{c|}{App. 1 (URLLC) - Un Coded} & \multicolumn{3}{c||}{App. 2 (eMBB) - Un Coded} & \multicolumn{3}{c|}{App. 1 - (URLLC) Coded} & \multicolumn{3}{c|}{App. 2 (eMBB) - Un Coded} \\ \hline
        
    ($|\mathcal{P}_1|, |\mathcal{P}_2|$) & ${E[D(\mathcal{P}_1)]}$ & ${E[I(\mathcal{P}_1)]}$ & ${E[G(\mathcal{P}_1)]}$ & ${E[D(\mathcal{P}_2)]}$ & ${E[I(\mathcal{P}_2)]}$ & ${E[G(\mathcal{P}_2)]}$ & ${E[D(\mathcal{P}_1)]}$ & ${E[I(\mathcal{P}_1)]}$ & ${E[G(\mathcal{P}_1)]}$ & ${E[D(\mathcal{P}_2)]}$ & ${E[I(\mathcal{P}_2)]}$ & ${E[G(\mathcal{P}_2)]}$ \\ \hline
    (5, 15) & 112.75 & 606.35 & 4.00 & 112.98 & 718.53 & \textcolor{purple}{\textbf{12.00}} & \textcolor{blue}{\textbf{94.39}} & 123.63 & 3.61 & 112.69 & 694.37 & \textcolor{purple}{\textbf{12.00}} \\ \hline
    (6, 14) & 112.94 & 641.05 & 4.80 & 112.76 & 721.35 & \textcolor{purple}{\textbf{11.21}} & \textcolor{blue}{\textbf{93.12}} & 130.14 & 4.33 & 112.78 & 700.82 & \textcolor{purple}{\textbf{11.16}} \\ \hline
    (7, 13) & 112.67 & 649.58 & 5.60 & 112.65 & 682.82 & \textcolor{purple}{\textbf{10.39}} & \textcolor{blue}{\textbf{91.18}} & 131.68 & 5.06 & 112.56 & 703.31 & \textcolor{purple}{\textbf{10.39}} \\ \hline
    (8, 12) & 112.63 & 651.93 & 6.41 & 112.72 & 692.53 & \textcolor{purple}{\textbf{9.60}} & \textcolor{blue}{\textbf{90.08}} & 136.19 & 5.78 & 112.84 & 694.10 & \textcolor{purple}{\textbf{9.59}} \\ \hline
    (9, 11) & 112.88 & 677.41 & 7.19 & 112.64 & 683.53 & 8.80 & \textcolor{blue}{\textbf{88.50}} & 134.99 & 6.51 & 112.72 & 681.09 & 8.80 \\ \hline
    (10, 10) & 112.86 & 686.04 & 8.00 & 112.67 & 670.10 & 7.98 & \textcolor{blue}{\textbf{88.57}} & 141.96 & 7.23 & 112.81 & 670.14 & 7.98 \\ \hline
    (11, 9) & 112.64 & 681.09 & 8.80 & 112.72 & 668.09 & 7.20 & \textcolor{blue}{\textbf{87.92}} & 144.62 & 7.95 & 112.85 & 664.13 & 7.20 \\ \hline
    (12, 8) & 112.89 & 683.22 & 9.60 & 112.73 & 656.51 & 6.39 & \textcolor{blue}{\textbf{87.41}} & 146.63 & 8.68 & 112.76 & 662.45 & 6.39 \\ \hline
    (13, 7) & 112.71 & 693.10 & 10.41 & 112.72 & 642.27 & 5.59 & \textcolor{blue}{\textbf{86.92}} & 149.95 & 9.40 & 112.89 & 640.77 & 5.60 \\ \hline
    (14, 6) & 112.83 & 701.44 & 11.22 & 112.60 & 635.95 & 4.81 & \textcolor{blue}{\textbf{87.11}} & 154.38 & 10.13 & 112.70 & 624.61 & 4.80 \\ \hline
    (15, 5) & 112.75 & 714.11 & 11.97 & 112.92 & 617.11 & 4.00 & \textcolor{blue}{\textbf{86.75}} & 157.66 & 10.85 & 112.80 & 608.91 & 3.99 \\ \hline
    \end{tabular}
    \vspace{-0.6cm}
\end{table*}

\section{Conclusions and Future Work}\label{sec:conc}

This work expands the benefits of network slicing in 5G networks and beyond, and presents a novel hybrid approach that incorporates coding into a fraction of virtual network slices. To rigorously evaluate the efficacy of our new hybrid approach, we considered performance requirements of the key 5G use cases, i.e., eMBB and URLLC, as identified by 3GPP. Although we considered a hybrid solution where coding is available only on a fraction of the slices in the network, how to optimally determine the amount of physical resources needed for the different applications and use cases in the entire network remains as future work. As an interesting envisioning technology, we propose combining the idea of resources allocation optimization and network sliced management, e.g., using SDN network controller \cite{mamushiane2018comparative,amin2018hybrid,cohen2021bringing} and using advanced coding solutions \cite{cohen2020adaptive,cohen2022broadcast}. The configuration of the virtual sliced network can therefore be self-organized. 
\vspace{-0.2cm}
\bibliographystyle{IEEEtran}
\bibliography{references}

\end{document}